\documentstyle [prl,aps,floats]{revtex}

\begin{document}

%\twocolumn[\hsize\textwidth\columnwidth\hsize\csname @twocolumnfalse\endcsname
\title{Kaluza-Kelin Higher Derivative Gravity and Friedmann-Robertson-Walker Cosmology}
\author{ W.F. Kao\thanks{email:wfgore@cc.nctu.edu.tw},
}
\address{ Institute of Physics, Chiao Tung
University, Hsinchu, Taiwan\\
}
\date\today
\maketitle

% Abstract
\begin{abstract} 
Kaluza-Klein approach in an $N(=1+3+D$)-dimensional Friedmann-Robertson-Walker type space is often adopted in the literature. 
We derive a compact expression for the Friedmann equation in a ($1+3+D$)-dimensional space.  
The redundancy of the associated field equations due to the Bianchi identity is analyzed. 
We also study the dilaton gravity theory with higher-derivative gravitational couplings. 
It turns out that higher-order terms will not affect the Friedmann equation in a constant flat internal space. 
This is true only for the flat-De Sitter external space.
The inflationary solution in an induced-gravity model is also discussed as an application.  

\end{abstract}
\vskip .2in

PACS numbers: 98.80Cq; 04.20 -q; 04.50.+h; 04.20 Cv
\section{Introduction}
Higher derivative terms should be important for the physics near the Planck scale \cite{kim,dm95}. 
For example, higher-order corrections derived from the quantum gravity or the string theory \cite{green} have been considered in the study of the inflationary universe \cite{acc,acc1}.
Higher derivative terms also arise as quantum corrections to matter fields \cite{green}.
Moreover, the stability analysis of pure higher-derivative models was shown in Ref. \cite{dm95}. 
It is hence interesting to extend this stability analysis to different models. 

On the other hand, Kaluza-Klein theory \cite{kk,visser} is also important for the evolution of the early universe. 
This is because that the dimensional-reduction process could affect the process of inflation significantly.
We are going to show that a constant internal-space solution to the induced Kaluza-Klein theory will put two additional constraints to the inflationary solution.

Note that, in four-dimensional space, $
E = R_{\lambda\mu\nu\rho}\,R^{\lambda\mu\nu\rho} -
4\,R_{\mu\nu}\,R^{\mu\nu}+ R^2
$ is the integrand of the Gauss-Bonnet term \cite{weinberg1}. 
In addition, Weyl tensor \cite{weinberg1} connects these fourth-order curvature terms in four-dimensional space.
Hence one only needs to deal with $R^2$ term in four-dimensional space.
This is not true in higher dimensional spaces. 
One would have to deal with all three different fourth-order terms. 
Hence equations of motion for the higher-derivative Kaluza-Klein theory are much more complicated than the four-dimensional higher-derivative gravity.
The method developed here will help reducing the complication due to the difficulty mentioned above.

We will derive a compact expression for the Friedmann equation \cite{kao99} in a ($1+3+D$)-dimensional space in this paper.  
The redundancy of the associated field equations due to the Bianchi Identity will also be analyzed. 
We will also study the induced-gravity theory \cite{zee,ni,kao00} with higher-derivative gravitational couplings. 
It turns out that higher-order terms will not affect the Friedmann equation in a constant flat internal space with a flat-De Sitter external space.
The inflationary solution in an induced-gravity model will also be discussed as an application \cite{zee}. 
All dimensionful parameters are replaced by scalar field couplings in induced-gravity models.
The values of these parameters measured today are interpreted as the vacuum expectation values of these scalar field couplings.
These induced models have also been a focus of research interest for a long time.

This paper will be organized as follows: (i) In section II, we will derive a model-independent expression for the Friedmann equation in higher-dimensional higher-derivative theory.
(ii) We will discuss the inflationary solution in a pure gravity theory in section III.
(iii) The induced-gravity theory will be discussed in section IV.
(iv) Finally, we will draw a few conclusions in section V.

\section{Friedmann Equation from FRW metric}

Note that, throughout this paper, the curvature tensor ${\bf R}^D_{ABC}
({\bf g}_{AB})$ will be defined by the following equation
\begin{equation}
[D_A,\,\, D_B]A_C = {\bf R}^{D}\,\,_{CBA} A_D .
\label{eqn:curv}
\end{equation}
Accordingly, one has
${\bf R}^{D}_{ABC} = - \partial_C {\bf \Gamma} ^D_{AB} -
{\bf \Gamma} ^E_{AB} {\bf \Gamma} ^D_{CE} - ( B \leftrightarrow C)$.
Here ${\bf \Gamma} ^C_{AB}$ denotes the Christoffel symbol (or spin connection of the covariant derivative, i.e. 
$D_A A_B \equiv \partial_A A_B - {\bf \Gamma} ^C_{AB}A_C$). 
To be more specific,
${\bf \Gamma}^C_{AB} = {1 \over 2} {\bf g}^{CD} (\partial_A
{\bf g}_{DB} + \partial_B {\bf g}_{DA} - \partial_D {\bf g}_{AB}
)$. 
Moreover, the Ricci tensor ${\bf R}_{AB}$ is defined as
\begin{equation}
{\bf R}_{AB} = {\bf R}^{C}\,\,_{ABC}. \label{ricci}
\end{equation}
And the scalar curvature ${\bf R}$ is defined as the trace of the Ricci tensor
${\bf R} \equiv {\bf g}^{AB} {\bf R}_{AB}$.
Also, one defines the Einstein tensor according to
${\bf G}_{AB} \equiv {1 \over 2} {\bf g}_{AB} {\bf R} - {\bf R}_{AB}$

Note that we will use bold-faced notation (e.g. ${\bf R}$) to denote field variable in $N(=4+D)$-dimensional space. 
In addition, normal notations will denote field variable evaluated in the $4$ or $D$-dimensional spaces as the compactification process $M^N \to M^4 \times M^D$ takes places.
Here $M^4$ is the four dimensional Friedmann-Robertson-Walker (FRW) space and $M^D$ is the compactified internal space. 
We will assume that $M^D$ is the $D$ dimensional FRW space for simplicity.
In fact the metric of $M^N$ is assumed to be 
\begin{eqnarray}
ds^2 &\equiv &{\bf g}_{AB} dZ^A dZ^B \equiv g_{\mu \nu} dx^\mu dx^\nu +
f_{mn}dz^mdz^n \\
 & =&  -b(t)^2dt^2
+ {a^2}(t) \Bigl( {dr^2  \over 1 - k_1 r^2} + r^2 d^3\Omega  \Bigr )
+ {d^2}(t) \Bigl( {dz^2  \over 1 - k_2 z^2} + z^2 d^D\Omega  \Bigr ) .
\label{eqn:frw} \label{GFRW}
\end{eqnarray}
Here $ d^p \Omega $ is the solid angle 
$d^p \Omega \, = \, d{\theta_1}^2 + {\sin}^2 \theta_1 \, d{\theta_2}^2 +
\cdots +\sin^2 \theta_1 \sin^2 \theta_2 \cdots \sin^2 \theta_{p-3}
d\theta_{p-2}^2$ and
$k_1,\,\, k_2 \, = \, 0, \pm 1$ stand for a flat, closed or open universe respectively.
Note that we will also write 
$g_{ij}=a^2 h_{ij}$ and $g_{mn} = d^2 h_{mn}$ for convenience.
Note also that this ($1+3+D$)-dimensional FRW metric can be obtained from the generalized Friedmann-Robertson-Walker (GFRW) (\ref{GFRW}) metric by setting the lapse function  $b(t)=1$ in Eq. (\ref{eqn:frw}).
Note that $\theta_i$ is the phase angle of the $D$-dimensional spherical coordinate. For example, one should write
\begin{equation}
z_1=z \sin \theta_1 \sin \theta_2 \cdots  \sin \theta_{D-2}.
\end{equation}
Note that we have used $N$-dimensional space-time coordinate as
$Z^{A} \to (x^\mu, \,\, z^m)$ for $A (= 0, 1, \cdots , N-1)$,
$\mu (= 0,1,2,3)$, and $m(=1,2, \cdots, D)$ denoting the $N$, $4$, and $D$-dimensional space-time indices. 
Therefore, capital Roman letters $A,B,C,
\cdots$ will denote $N$ dimensional indices. 
In addition, Greek letters will denote $4$-indices while the second half of the Roman letters will denote $D$-dimensional space-time indices. 
Here we have assumed that the internal space ($z$) is independent of the external space ($x$). 
The only dependence on $t$ is through the scale factor $d(t)$.

Therefore, one can show that
\begin{eqnarray}
{\bf \Gamma}^\gamma_{\mu \nu} &=& \Gamma^\gamma_{\mu \nu}  ,\\
{\bf \Gamma}^p_{mn} &=& \Gamma^p_{mn} , \\
{\bf \Gamma}^\gamma_{mn} &=& - \partial^\gamma \beta g_{mn} ,\\
{\bf \Gamma}^p_{\mu m} &=& \partial_\mu \beta \delta^p_m .
\end{eqnarray}
Here $\partial_\mu \beta \equiv {\partial_\mu d / d}$ with a non-vanishing $t$-component.
Hence, we will write $I = \partial_t \beta$ for convenience from now on.
One can also show that
\begin{eqnarray}
{\bf R}^{ti}_{\,\,\,\,tj}&=&{1\over 2} [H\dot{B}+2B(\dot{H}+H^2)]\delta^i_j ,
\label{Rti} \\
{\bf R}^{ij}_{\,\,\,\,kl}&=& (H^2B+{k_1 \over a^2} ) (\delta^i_k \delta^j_l
-\delta^i_l
\delta^j_k) \label{Rkl} , \\
{\bf R}^{tm}_{\,\,\,\,tn}&=&{1\over 2} [I\dot{B}+2B(\dot{I}+I^2)]\delta^m_n  ,
\label{Rtm}\\
{\bf R}^{im}_{\,\,\,\,jn}&=&-HI\delta^m_n \delta^i_j ,
\label{Rim}\\
{\bf R}^{mn}_{\,\,\,\,pq}&=& (I^2B+{k_2 \over d^2} ) (\delta^m_p \delta^n_q
-\delta^m_q \delta^n_p )\label{Rmn} .
\end{eqnarray}
Accordingly, one has
\begin{eqnarray}
{\bf R}^{t}_{\,\,t}&=& -3(\dot{H}+H^2) -D(\dot{I}+I^2)  ,
\label{Rtt}\\
{\bf R}^{i}_{\,\,j}&=&- [\dot{H}+3H^2+2{k_1 \over a^2}-DHI]\delta^i_j
\label{Rij}, \\
{\bf R}^{m}_{\,\,n}&=& -[\dot{I}+DI^2+(D-1){k_2 \over d^2}-3HI] \delta^m_n ,
\label{Rm} \\
{\bf R}&=& - [ 6(\dot{H}+2H^2+{k_1 \over
a^2}-DHI)+D(D-1)(I^2+{k_2 \over d^2}) +2D(\dot{I}+I^2)]
\label{R}
\end{eqnarray}
once we set $b=1$.

The variational equation of $b$ will give \cite{kao99}
\begin{equation}
L - H {\delta L \over \delta H} - I {\delta L \over \delta I}  +
[ H {d \over  dt}+ H( 3H +DI) -\dot{H} ] {\delta L \over \delta \dot{H}}
+ [ I {d \over  dt}+ I( 3H +DI) -\dot{I} ] {\delta L \over \delta \dot{I}}
\equiv {\cal D} L                            =0  \label{key}  .
\end{equation}
In addition, variational equations of $a$ and $d$ also give
\begin{eqnarray} \label{aeq}
3L- H {\delta L \over \delta H} + (H^2- \dot{H}) {\delta L \over \delta
\dot{H}} &&
-(2H +DI + {d \over  dt})  [ -(4H +DI + {d \over  dt})
{\delta L \over \delta \dot{H}}  +{\delta L \over \delta {H}} ]
-2k_1 {\delta L \over \delta k_1} \equiv {\cal D}_3 L =0  ,
\\ \label{deq}
DL- I {\delta L \over \delta I} + (I^2- \dot{I}) {\delta L \over \delta
\dot{I}} && -
(3H +(D-1)I + {d \over  dt})  [ -(3H +(D+1)I + {d \over  dt})
{\delta L \over \delta \dot{I}}  +{\delta L \over \delta {I}} ]
-2k_2 {\delta L \over \delta k_2} \equiv {\cal D}_D L =0 .
\end{eqnarray}
Note also that the Bianchi identity $D_MG^{MN}=0$ can be brought to the following form:
\begin{equation}
(\partial_t + 3 H +DI) H_{tt} + 3 a^2H H_{3} +D d^2 I H_D=0. \label{h3}
\end{equation}
Here $H_{ij} = H_{n-1} h_{ij}$.
Alternatively, one can show that
\begin{equation}
(d/dt +3H +DI) {\cal D} L = H{\cal D}_3 L +I {\cal D}_D L \label{biN}
\end{equation}
from direct differentiating ${\cal D} L$. 
This in fact clarifies that
${\cal D} L \sim H_{tt}$, ${\cal D}_3 L \sim H_3$ and ${\cal D}_D L \sim
H_D$. 
Therefore, one can ignore either one of  $H_3$ and $H_D$ equation without losing any information. 
Moreover, this conclusion also indicates that both $H_3$ and $H_D$ are equally redundant. 
This is, however, not true under the condition where $d=$constant, or $I=0$. 
As indicated by Eq. (\ref{biN}), one has instead
\begin{equation}
(d/dt +3H) {\cal D} \bar{L} = H{\cal D}_3 \bar{L}  \label{bi3}
\end{equation}
under this condition.
Here we have written ${\cal D} \bar{L} \equiv {\cal D} L |_{I=0}$ and similarly for
${\cal D}_3 \bar{L}$,  ${\cal D}_D \bar{L}$, $\bar{L}$, etc.
Therefore, the reduced Bianchi Identity (\ref{bi3}) only tells us that
${\cal D}_3 \bar{L}=0$ is redundant to the Friedman equation 
${\cal D} \bar{L}=0$ under the constraint $I=0$. 
In another word, one still has to consider the left-over $d$-equation ${\cal D}_D \bar{L}=0$ in order to solve the constant internal-space solution. 
This point is often overlooked and should be carefully addressed in studying constant internal-space solutions.

To be more explicitly, the Friedmann equation (\ref{key}) and the $a(t)$ equation (\ref{aeq}) become
\begin{eqnarray}
\bar{L} - H {\delta \bar{L} \over \delta H}  &+&
[ H {d \over  dt}+ 3H^2  -\dot{H} ] {\delta \bar{L} \over \delta \dot{H}}
                            =0  \label{key4} ,
\\ \label{aeq4}
3\bar{L}- H {\delta \bar{L} \over \delta H} + (H^2- \dot{H}) {\delta \bar{L} \over \delta
\dot{H}} &=&
(2H + {d \over  dt})  [ -(4H + {d \over  dt})
{\delta \bar{L} \over \delta \dot{H}}  +{\delta \bar{L} \over \delta {H}} ]
+2k_1 {\delta \bar{L} \over \delta k_1}  , \\
\label{deq4}
D\bar{L} &=&
(3H + {d \over  dt})  [ -(3H + {d \over  dt})
{\delta \bar{L} \over \delta \dot{I}}  +{\delta \bar{L} \over \delta {I}} ]
+2k_2 {\delta \bar{L} \over \delta k_2} 
\end{eqnarray}
upon assuming $d(t)=$ constant.
Therefore, one should try to solve the system directly from analyzing both Eq.s
(\ref{key4}) and (\ref{deq4}).

\section{Higher Derivative Gravity}
We will be working on the solution with constant internal space from now on for simplicity.
This kind of solution has been a focus of study in the literature. 
Part of the reason is probably due to the increasing complexity in the Kaluza-Klein theory. 
It also applies to models that compactification is assumed to have been completed well before inflation.
As a result, all equations are replaced by reduced barred equations shown earlier.
One has to, however, pay attention to the left over internal $I$-equation serving as an additional constraint to field equations.

In the derivation of the field equation, we will need a complete list of all squared-curvature terms:
\begin{eqnarray}
\label{R42}
({\bf R}^{AB}_{\;\;CD})^2 &=&
12 (\dot{H} +H^2)^2 +4D (\dot{I} +I^2)^2 + 12DH^2I^2
+12 (H^2 + {k_1 \over a^2})^2 +2D(D-1)(I^2 + {k_2 \over d^2})^2  ,\\
\label{R22}
({\bf R}^A_{\;B})^2 &=&
12 (\dot{H} +H^2)^2 +D(D+1) (\dot{I} +I^2)^2
+12 (H^2 + {k_1 \over a^2})^2 +D(D-1)^2(I^2 +  {k_2 \over d^2})^2
+ 3D(D+3)H^2I^2
\nonumber \\
&+& 12 (\dot{H} +H^2)
(H^2 + {k_1 \over a^2})
+6D(\dot{H}+H^2 -HI)(\dot{I}+I^2)
-6DHI (\dot{H}+3H^2+2{k_1 \over a^2}) \nonumber \\
&+& 2D(D-1)(\dot{I}+ I^2)(I^2 + {k_2 \over d^2})
-6D(D-1)HI(I^2 + {k_2 \over d^2}) ,
\\
\label{R02}
({\bf R})^2 &=&
36 (\dot{H} +H^2)^2 +4D^2 (\dot{I} +I^2)^2 + 36D^2H^2I^2
+36 (H^2 + {k_1 \over a^2})^2 +D^2(D-1)^2(I^2 + {k_2 \over d^2})^2
\nonumber \\
&+& 72 (\dot{H} +H^2)
(H^2 + {k_1 \over a^2}-DHI) +12D(D-1)(\dot{H}+2H^2+ {k_1 \over
a^2} )(I^2 + {k_2 \over d^2})
\nonumber \\
&+&24D(\dot{H}+2H^2+{k_1 \over a^2})(\dot{I}+I^2)
-72DHI (H^2+{k_1 \over a^2})
+4D^2(D-1)(\dot{I}+ I^2)(I^2 + {k_2 \over d^2})
\nonumber \\
&-& 24D^2(\dot{I}+ I^2)HI
-12D^2(D-1)HI(I^2 + {k_2 \over d^2}) .
\end{eqnarray}
Let us consider the system described by the pure-gravity effective Lagrangian
\begin{equation}
L=-{\bf R} -c_1 ({\bf R}^{AB}_{\;\;CD})^2 -
c_2 ({\bf R}^{A}_{\;B})^2 -
c_3 {\bf R}^2.
\end{equation}
One can show that
\begin{eqnarray}
{\delta \over \delta I} \bar{L} &=&
-6DH \left \{ 1-c_2  \left [ \dot{H} +3H^2 +2 {k_1 \over a^2} + {\Lambda \over D} \right ]
-12c_3 \left [ \dot{H} +2H^2 + {k_1 \over a^2} +{\Lambda \over 6}  \right ]
\right \}  \label{frd3}  , \\
{\delta \over \delta \dot{I}} \bar{L} &=&
2D \left \{ 1-c_2 \left [ 3\dot{H} +3H^2 +  {\Lambda \over D} \right  ]
-12c_3 \left [ \dot{H} +2H^2 + {k_1 \over a^2} + {\Lambda \over 6} \right ] \right \}.
\label{deqd}
\end{eqnarray}
In addition, one has
\begin{equation} \label{k2eq}
k_2 {\delta \over \delta k_2} \bar{L} = \Lambda -\Lambda^2 \left [
{4c_1 \over D(D-1)} +{2c_2 \over D} +2c_3 \right ] -12c_3 \Lambda  [(\dot{H} +2H^2
+{k_1 \over a^2}) ] .
\end{equation}
Here we have defined $\Lambda \equiv D(D-1)k_2 /d^2$.
In addition, one can show that
\begin{eqnarray}
\bar{L}&=& 6 [ \dot{H} +2H^2 +{k_1 \over a^2} ]
-12 (c_1+c_2+3 c_3) [ (\dot{H} +H^2)^2 + (H^2 +{k_1 \over a^2})^2]
-12 (c_2+6 c_3) [ (\dot{H} +H^2)(H^2 +{k_1 \over a^2})]
\nonumber \\
&& -12 c_3 \Lambda [ \dot{H} +2 H^2 +{k_1 \over a^2}]
+ \Lambda - [2c_1/D(D-1)+c_2/D+c_3]\Lambda^2 . \label{barL}
\end{eqnarray}
Hence one can show further that
\begin{eqnarray}
{\delta \over \delta \dot{H} } \bar{L} &=&
6 - 24 (c_1 +c_2 +3c_3) (\dot{H} +H^2)  - 12 (c_2 +6c_3) ( H^2 + {k_1 \over a^2} )
- 12c_3 \Lambda  , \\
{\delta \over \delta H} \bar{L} &=& 24 H- 24(2c_1+3c_2+12c_3)H
\left [ \dot{H} +2H^2 + {k_1 \over a^2}  \right ] -
48c_3H \Lambda .
\end{eqnarray}

Before we press on, we will prove a theorem concerning the role played by these $R^2$-terms in the case when $k_1=k_2=0$. 
Note that recent measurements \cite{k=0} indicate that we live in a flat 3-space to a very high precision. 
We will hence assume that $k_1=0$ from now on. 
Note also that flat internal space solutions have attracted lots of attention lately \cite{visser}.

{\bf Theorem} All three different fourth-derivative $R^2$-type terms in the Lagrangian (\ref{barL}) can be removed from $\bar{L}$ without affecting the Friedmann equation (\ref{key4}) when one considers the expanding De Sitter solution ($H=H_0$) under the case that $k_1=k_2=0$. Here we have also assumed that we are working with constant internal-space solutions.

Note that this theorem states that fourth-derivative terms can not affect the $4$-dimensional Friedmann equation for all $4$-dimensional flat-De Sitter models where $H=H_0$.
Above theorem remains true in the higher dimensional model provided that there is no internal-space contribution to the Friedmann equation.
This will be the point we will follow for the proof.  
This theorem can be proved as follows: Note that under flat-De Sitter conditions:
(i) $H=H_0$, (ii) $k_1=0$, and (iii) $k_2=0$ (in fact only $\Lambda =0$ is required), one can write 
$A \equiv \dot{H} +H^2=H_0^2$ and $B\equiv H^2+k_1/a^2=H_0^2$.
Hence one can show that
$\bar{L}_2=\bar{L}_2 (A^2+B^2)$ with $\bar{L}_2$ denoting fourth-derivative terms.
This is true because the term proportional to
$AB$ can be shown to be a total derivative when the metric measure $\sqrt{g}$ is  incorporated, namely, one can show that 
\begin{equation}
a^3 AB =d \,[\dot{a}^3/3 +k_1\dot{a} ] /dt.
\end{equation}
Therefore, the $AB$-term is indeed a surface term. 
Hence it can be ignored even conditions 
$H=H_0$ and $k_1=\Lambda=0$ do not apply. 
Hence, one can show that
\begin{eqnarray}
\delta \bar{L}_2 /\delta \dot{H} &=& 2A \delta \bar{L}_2 /\delta (A^2+B^2) =
2H_0^2 \delta \bar{L}_2 /\delta (A^2+B^2) , \\
\delta \bar{L}_2 /\delta H^2 &=& 2(A+B) \delta \bar{L}_2 /\delta (A^2+B^2) =
4H_0^2 \delta \bar{L}_2 /\delta (A^2+B^2)=2\delta \bar{L}_2 /\delta \dot{H}
\end{eqnarray}
under flat-De Sitter conditions.
One can further cast the Friedmann equation as
\begin{equation}
\bar{L}_2 =
H_0 {\delta \bar{L}_2 \over \delta H}
-3H_0^2 {\delta \bar{L}_2 \over \delta \dot{H}} 
={1 \over 2} H_0^2 {\delta \bar{L}_2 \over \delta H_0^2}
\end{equation}
under the flat-De Sitter conditions.
By using the fact that $\bar{L}_2  \propto H_0^4$ under flat-De Sitter conditions, one can directly show that the above fourth-derivative part of the Friedmann equation is nothing more than a simple identity. 
Therefore, one shows that one can freely remove the contribution of the fourth-derivative terms from the effective Lagrangian without affecting the flat-De Sitter space Friedmann equation (\ref{key4}). 

One can easily show that, from the Friedmann equation (\ref{key4}), the pure-gravity theory given by the effective Lagrangian (\ref{barL}) can not support a non-zero $H_0$ solution for the reason similar to the quadratic-Einstein theory. 
Therefore, one has to consider the case that $k_2 \ne 0$.
As a result, an effective cosmological-constant term may be present to support an expanding-De Sitter phase.

Indeed, one can show that
\begin{equation}
H_0^2 = - {\Lambda \left[ 1- \left(
{2c_1 \over D(D-1)} +{c_2 \over D} +c_3 \right) \Lambda \right] 
\over 
6- 12c_3 \Lambda} \label{ph1}
\end{equation}
from the Friedmann equation.
On the other hand, there is another constraint (\ref{deq4}) left over from the constant internal-space assumption. 
This will give instead another equation for $H_0$:
\begin{equation} \label{ph2}
2D(c_1 +6c_2 +42c_3)H_0^4 - \{ 4D- [3c_2+ 4(2D-1)c_3] \Lambda \} H_0^2
-{1 \over 12} \left [ (D-2) \Lambda - (D-4) \left(
{2c_1 \over D(D-1)} +{c_2 \over D} +c_3 \right)  \Lambda^2 \right ]=0.
\end{equation}
Eq.s (\ref{ph1}-\ref{ph2}) give a set of two algebraic equations for the system.
Hence field parameters can easily be adjusted to get solution one is looking for.

\section{Dilaton Gravity}
We will study the inflationary universe for dilaton-gravity models. 
For the moment, let us consider the pure dilaton gravity model given by the effective Lagrangian
\begin{equation}
L_s \equiv L +T_\phi -V(\phi) \label{pdg}
\end{equation}
with $T_\phi \equiv - (\partial \phi)^2/2$.
Here Lagrangian $L$ denotes those parts independent of the dilaton field.
We will compare later its result with the induced gravity model given instead by
\begin{equation}
L_{{\rm id}}=L+L_\phi \equiv L_1+L_2+L_\phi \label{idg}
\end{equation}
with $L_1 \equiv - \epsilon  \phi^2 {\bf R} /2$ denoting the induced term and
$L_2$ denoting higher-derivative terms. 
Here $L_\phi=T_\phi-V(\phi)$ denotes the scalar field term with an arbitrary potential to be determined from the equation of motion.
Similar to earlier discussions, we will assume that $I=0$ in this section too.
One can show that the Friedmann equation for the dilaton gravity system (\ref{pdg}) reads:
\begin{equation}
\bar{L} - H {\delta \bar{L} \over \delta H}  +
[ H {d \over  dt}+ 3H^2  -\dot{H} ] {\delta \bar{L} \over \delta \dot{H}}
                            =T_\phi + V(\phi)  \label{key4phi} .
\end{equation}
One can easily show that the leading-order inflationary solution gives 
\begin{equation}
H_0^2 =V_0 /6= 7/4(c_1 +6c_2+42c_3)
\end{equation}
under the slow-rollover condition $(\dot{\phi}/\phi)^2 \ll H_0^2$. 
Therefore, one finds that the presence of the higher-derivative term and the static internal-space condition impose very strong constraints on possible coupled field parameters $c_i$ in order for an inflating external $4$-space to exist.

Next, we will study the induced-gravity model given by expression (\ref{idg}).
One can show that the Friedmann equation reads
\begin{equation}
{1 \over 2} \dot{\phi}^2 +V = 3 \epsilon \phi^2H^2 +6 \epsilon H \phi \dot{\phi} +K
\label{f-eqid}
\end{equation}
with $K$ denoting terms derived from the higher-order contribution.
Explicitly, 
\begin{equation}
K=12(c_1+c_2+3c_3)(\dot{H}^2 - 2H \ddot{H} - 6H^2 \dot{H}).
\end{equation}
Furthermore, the $\phi$-equation can be shown to be
\begin{equation}
6 \epsilon \phi (\dot{H} + 2H^2) = \ddot{\phi} +3 H \dot{\phi} +V' .
\end{equation}
In addition, one can derive the following scalar equation from the Bianchi combination of the Friedmann equation. 
It reads
\begin{equation}
\ddot{\phi} +3 H\dot{\phi} + {\dot{\phi}^2 \over \phi} =
{1 \over (1+6 \epsilon) \phi} [4V-\phi V'] -  
{1 \over (1+6 \epsilon) H\phi} [\dot{K} +4HK]  . \label{ddphi}
\end{equation}
On setting $k_1=k_2=0$, $\phi=\phi_0$, $I=0$, and  $H=H_0+\delta H$, one can show that the leading-order Friedmann equation (\ref{f-eqid}) gives
\begin{equation}
V_0= 3 \epsilon \phi_0^2H_0^2.
\end{equation} 
Moreover, the scalar equation gives 
\begin{equation}
12\epsilon \phi_0H_0^2=V_0'.
\end{equation}
Here we have assumed the slow-rollover approximation 
$|\dot{\phi}/\phi|  \ll H_0$, and $|\ddot{\phi}/\phi| \ll H_0^2 $.
Hence these leading-order perturbation equations show that some field parameters are related to each other via
\begin{equation} \label{constraint}
4V_0 =\phi_0 {\partial V \over \partial \phi} (\phi= \phi_0) .
\end{equation}
This condition is referred to as scaling condition in Ref. \cite{kao00}.

In addition, the first-order perturbation equation of the Friedmann equation (\ref{f-eqid}) gives
\begin{equation}
H_0^2 = {\epsilon \phi_0^2 \over 16(c_1+c_2 +3c_3)}. \label{h31}
\end{equation}
The leading-order perturbation equation of $\phi$-equation (\ref{ddphi}) gives
\begin{equation}
\delta H = \exp [-4H_0t]. \label{dh} 
\end{equation}
In addition, the leading-order and first-order perturbation $I$-equations give
\begin{eqnarray}
H_0^2 &=& {7 \epsilon \phi_0^2 \over 8 (c_1+6c_2+42c_3) } , \label{h1} \\
H_0^2 &=& {\epsilon \phi_0^2 \over (4c_2+48c_3) } \label{h2}
\end{eqnarray}
respectively. 
Note that the $I$-equation in the presence of a slow-rollover scalar field potential is
\begin{equation} \label{Ieqi}
D\bar{L} =
(3H + {d \over  dt})  [ -(3H + {d \over  dt})
{\delta \bar{L} \over \delta \dot{I}}  +{\delta \bar{L} \over \delta {I}} ]
+DV_0.
\end{equation}
Note also that, in deriving above equations, we have used the solution to $\delta H$ given by Eq. (\ref{dh}).
Here the effective Lagrangian $\bar{L}$ is
\begin{equation}
\bar{L} = 3 \epsilon \phi_0^2  [ \dot{H} +2H^2 ]
-12 (c_1+c_2+3 c_3) [ (\dot{H} +H^2)^2 + H^4]
-12 (c_2+6 c_3) [ (\dot{H} +H^2)H^2 ] \label{barLi}
\end{equation}
under the slow-rollover approximation and the conditions $k_1=k_2=0$.
In addition, the variations of the $I$-equation are
\begin{eqnarray}
{\delta \over \delta I} \bar{L} &=&
-6DH \left \{ {\epsilon \over 2} \phi_0^2 -c_2  \left [ \dot{H} +3H^2   \right ]
-12c_3 \left [ \dot{H} +2H^2  \right ]
\right \}  \label{frd3i}  , \\
{\delta \over \delta \dot{I}} \bar{L} &=&
2D \left \{ {\epsilon \over 2} \phi_0^2-c_2 \left [ 3\dot{H} +3H^2  \right  ]
-12c_3 \left [ \dot{H} +2H^2  \right ] \right \}
\label{deqdi}
\end{eqnarray}
in the same limit.

These equations impose strong constraints on field parameters that could admit an inflationary solution.
Therefore, one shows that the consistency of the perturbation equations gives us three expressions (\ref{h31},\ref{h1},\ref{h2}) for $H_0^2$. 
These stationary equations can be used to show that $c_1=c_2=0$.
%\begin{eqnarray}
%c_1 &=& 42c_3/55 \\
%c_2 &=& -96c_3/55.
%\end{eqnarray}
Hence, one has
\begin{equation}
H_0^2= { \epsilon \phi_0^2 \over 48 c_3}. \label{heq}
\end{equation}
Therefore, all the conclusion drawn in Ref. \cite{kao00} still applied here.

For example, we can consider the following effective symmetry-breaking potential \cite{kao00}
\begin{equation}
V={\lambda_1 \over 4} (\phi^2 - v^2)^2 + {\lambda_2 \over 4} \phi^4 -\Lambda_0.
\label{effV}
\end{equation}
Eq. (\ref{constraint}) and (\ref{heq}) show that this potential should take the following form:
\begin{equation}
V= {\lambda \over 4} \phi^4
- ( {\lambda \over 2} - { \epsilon^2 \over 32 c_3} ) \phi_0^2 \phi^2
+ ({\lambda  \over 4} - { \epsilon^2 \over 64 c_3} ) \phi_0^4 \label{Veff}.
\end{equation}
Here $\lambda\equiv\lambda_1 + \lambda_2$.
This is the form of the most general extended $\phi^4$ SSB potential that could admit a stable inflationary solution. 

The minimum of this potential is 
\begin{equation}
V_m \equiv V(\phi_m)= (  \epsilon^2  / 16 c_3 )
[ 1/4  -  \epsilon^2 /  64 c_3 \lambda  ]\phi_0^4= (  \epsilon^2 / 16 c_3 \lambda)V(0)
\end{equation}
when $\phi^2=\phi_m^2 \equiv (1-   \epsilon^2 /16 c_3 \lambda )\phi_0^2$.
Here $V(0)\equiv V(\phi=0)$ is the maximum of $V$.
One can also show that $0<V_m < V(0)$ provided that $c_3 >0$.
This model indicates that we will end up with a rather big cosmological constant of the order $10^{-6}$ if the extended $\phi^4$ model is in effect.
Note that the effective gravitational constant and cosmological constant observed in the post-inflationary phase are $1/4\pi G=\epsilon \phi_m^2=(1- \epsilon^2 /16 c_3 \lambda ) \epsilon \phi_0^2$
and $V_m = 3H_0^2/2$ respectively.
Here we have set $\epsilon \phi_m^2/2=1$ in Planck unit.
If the scale factor $a(t)$ is capable of expanding some $60$ e-fold
in a time interval of roughly $\Delta T \sim 10^8$ Planck unit, the Hubble constant should be of the order $H_0^2 \sim 10^{-6}$ in Planck unit.
Therefore, one has a rather big cosmological constant after inflation.

If we consider the model given by the following symmetry-breaking Coleman-Weinberg potential from radiative correction \cite{coleman}
\begin{equation}
V_{\rm c}={\lambda_3 \over 4} \phi^4 \ln ({\phi \over  w})^4 + {\lambda_4 \over 4}
\phi^4 - \Lambda_1,
\label{effVc}
\end{equation}
 one can show that the consistent potential takes the following form:
\begin{equation}
V_{\rm c}={\lambda_3 \over 4} \phi^4 \ln ({\phi \over  \phi_0})^4 + { \epsilon^2
\over  64 c_3} \phi^4 - {\lambda_3 \over 4} (\phi^4-
\phi_0^4).
\label{effVcm}
\end{equation}
The minimum of this potential is 
\begin{equation}
V_{{\rm c}m}=(\lambda_1 \phi_0^4/4) \{ 1- \exp [-\epsilon^2 /16 c_3
\lambda_1] \}
\end{equation}
when $\phi=\phi_m = \phi_0 \exp [-\epsilon^2 /64 c_3 \lambda_1]$.
Therefore, one can show that this model also give a rather big cosmological constant after inflation.
This may have to do with the field contents of the early universe \cite{weinberg}.
Hopefully, the soft-expansion era will be dominated by another lower-order induced-gravity model \cite{smolin}. 
Therefore, the re-heating process will be taken over by that lower-order effective induced-gravity model \cite{acc}.
It is still true for a small higher-order correction, namely, $c_3 \ll 1$.

\section{Conclusion}
In summary, the presence of a constant internal space puts two additional constraints on the parameters $c_i$ for the induced-gravity model.
These constraints are derived from stability conditions of the leading-order and first-order perturbation equations.
Therefore, one finds that the coefficients $c_1=c_2=0$ with only one degree of freedom left over.
The result is that the higher dimensional contribution of a constant flat internal space makes an interesting modification to the external-space equation.
This makes the Kaluza-Klein theory, at the lower-energy limit, behave rather similar to the induced higher derivative gravity shown in Ref. \cite{kao00}. 

{\bf Acknowledgments :}
This work is supported in part by the National Science Council under
the contract number NSC88-2112-M009-001.

%%%%%%%%%%%%%%%%%%%%%%%%%%%%%%%%%%%%%%%%%%%%%%%%%%%%%%%%%%%%%%%%%%%%%%%%%

\end{document}